\documentclass[12pt, reqno]{amsart}

\usepackage[a4paper, margin=1.in, foot=.3in]{geometry}
\usepackage[a4paper, unicode, colorlinks, pdfstartview=FitH, draft=false]{hyperref}
\usepackage[noBBpl,slantedGreek]{mathpazo}

\allowdisplaybreaks

\numberwithin{equation}{section}

\newtheorem{conjecture}{Conjecture}[section]
\newtheorem{corollary}{Corollary}[section]

\DeclareMathOperator{\cn}{cn}
\DeclareMathOperator{\dn}{dn}
\DeclareMathOperator{\sn}{sn}

\def \calH {\mathcal H}
\def \calI {\mathcal I}
\def \calP {\mathcal P}
\def \calR {\mathcal R}
\def \calS {\mathcal S}
\def \calT {\mathcal T}
\def \calU {\mathcal U}

\def \rmi {\mathrm i}

\begin{document}

\title{A possible combinatorial point for XYZ-spin chain}

\author{A. V. Razumov}
\address{Institute for High Energy Physics, 142281 Protvino, Moscow region, Russia}
\email{Alexander.Razumov@ihep.ru}

\author{Yu. G. Stroganov}
\address{Institute for High Energy Physics, 142281 Protvino, Moscow region, Russia}
\email{Yuri.Stroganov@ihep.ru}

\begin{abstract}
We formulate and discuss a number of conjectures on the ground state vectors of the XYZ-spin chains of odd length with periodic boundary conditions and a special choice of the Hamiltonian parameters. In particular, arguments for the validity of a sum rule for the components, which describes in a sense the degree of antiferromagneticity of the chain, are given.
\end{abstract}

\maketitle

\tableofcontents

\section{Introduction}

The two-dimensional eight-vertex lattice model is closely related to the quantum one-dimensional XYZ-spin chain. As far as we know, for the first time this was noticed by Sutherland in the paper \cite{Sut70}, where it was shown that the Hamiltonian of the periodic XYZ-spin chain
\begin{equation}
\calH_{\mathrm{XYZ}}= - \frac{1}{2} \sum_{j=1}^{N} [J_x \sigma_j^x\sigma_{j+1}^x + J_y \sigma_j^y \sigma_{j+1}^y + J_z \sigma_j^z\sigma_{j+1}^z]
\label{H2}
\end{equation}
commutes with the transfer-matrix of the periodic eight-vertex model if the weights of the latter are chosen in a special way. In distinction to the six-vertex model and the related to it quantum XXZ-spin chain, which is also described by the Hamiltonian (\ref{H2}) with the condition $J_x = J_y$, the eight-vertex model and the XYZ-spin chain cannot be solved via the Bethe ansatz. In 1971, Baxter proposed and used for investigation of these models an alternative method with the help of which he succeeded to find the partition function of the eight-vertex model in the thermodynamic limit~\cite{Bax71a, Bax72a}. In the same year he established a connection of the Hamiltonian (\ref{H2}) and the logarithmic derivative of the commuting family of transfer-matrices of the eight-vertex model over the spectral parameter, and found the ground state energy per site of the XYZ-spin chain in the limit of infinite number of sites~\cite{Bax71b, Bax72b}.

In the paper~\cite{Bax72b}, Baxter remarked also that the ground state energy of the XYZ-spin chain per site in the limit of infinite number of sites has a simple form
\begin{equation}
\lim_{N \rightarrow \infty} \frac{E}{N} = - \frac{1}{2}(J_x + J_y + J_z),
\label{infty}
\end{equation}
if the parameters $J_x$, $J_y$ and $J_z$ satisfy the relation
\begin{equation}
\label{condition}
J_x J_y + J_y J_z + J_z J_x = 0
\end{equation}
and belong to the domain
\begin{equation}
J_x + J_y + J_z > 0.
\label{condition2}
\end{equation}
Later, he established that the inversion relations for the eigenvalues of the commuting family of the transfer-matrices of the eight-vertex model have in this case a very simple solution even for the chains of finite length~\cite{Bax89}. Note that not for any solution of the inversion relations there is a corresponding eigenvector. However, if an eigenvector corresponding to the solution found by Baxter exists, it is an eigenvector of the Hamiltonian (\ref{H2}) with the eigenvalue
\begin{equation}
E = - \frac{N}{2} \, (J_x + J_y + J_z),  \label{sev}
\end{equation}
which is in the remarkable agreement with the formula (\ref{infty}).

As was noticed by one of the authors of the present paper~\cite{Str01a}, an eigenvector corresponding to the solution of the inversion relations found by Baxter exist for $N = 1, 3, 5, 7$, but for $N = 2, 4, 6$ there is now such a vector. This allowed to formulate a conjecture that the vector under consideration exists only for the chains of {\em odd\/} length and in this case it is a ground state of the system. The validity of this conjecture for the case of the XXZ-spin chain was established in the papers \cite{YanFen04, VenWoi06, RazStrZin07}. The corresponding explicit solution of the Baxter's $TQ$-equations also was found \cite{FriStrZag00, FriStrZag01}, that, in particular, allowed to find expressions for the simplest correlation functions \cite{Str01b, Str01cE}.

In the limit of the XXZ-spin chain we can without loss of generality choose the energy scale for which $J_x = J_y = 1$, then it follows from the relation (\ref{condition}) that $J_z = -1/2$. Here the matrix elements of the Hamiltonian (\ref{H2}) with respect to a natural basis, formed by the eigenvectors of the operators $\sigma^z_j$, are rational numbers, and the ground state energy is the rational number $-3 N/4$. It is clear that in this case we can normalize the ground state vector so that its components are integers. We found that for all odd $N \le 17$ the components are {\em positive\/} integers~\cite{RazStr01a}. It appeared also that some components and some sums of the components are hypothetically related to enumerations of the alternating-sign matrices~\cite{RazStr01a, RazStr06a}. Some of the conjectures formulated in the paper \cite{RazStr01a} have been proved already \cite{RazStr06a, KitMaiSlaTer02a, KitMaiSlaTer02b, DiFZinZub06}, and some of them have been generalized to the case of different boundary conditions \cite{BatdeGNie01, RazStr01b}.

The present paper is devoted to investigation of the XYZ-spin chain. Assuming that the conditions (\ref{condition}) and (\ref{condition2}) are satisfied, we found the explicit form of an eigenvector of the Hamiltonian $\calH_{\mathrm{XYZ}}$ with the eigenvalue (\ref{sev}) for all odd $N \le 19$. The obtained data allowed us to suggest a number of conjectures on the properties of the vector under consideration. In particular, as for the case of the XXZ-spin chain, we succeeded to trace a connection to combinatorial objects. We give also arguments for the validity of a sum rule for the components, which describes in a sense the degree of antiferromagneticity of the chain.

It would be interesting to find for the case under consideration the corresponding solution to the Baxter's $TQ$-equations. Till now we have not succeeded in it. A promising advance in this direction are the papers by Bazhanov and Mangazeev~\cite{BazMan05, BazMan06}, where, in particular, a recursive procedure to construct the solution is proposed.

\section{Eight-vertex model and XYZ-chain}

We use the parameterization of the Boltzmann weights of the eight-vertex model proposed by Baxter~\cite{Bax72a}. It has the form
\begin{align}
& a(v; \rho, \eta, k) = \rho \, \Theta (2\eta) \Theta (v - \eta) H(v + \eta), \label{BW1} \\
& b(v; \rho, \eta, k) = \rho \, \Theta (2\eta)  H(v - \eta)  \Theta (v + \eta), \label{BW2} \\
& c(v; \rho, \eta, k) = \rho \, H(2\eta) \Theta (v - \eta) \Theta (v + \eta), \label{BW3} \\
& d(v; \rho, \eta, k) = \rho \, H(2\eta) H(v - \eta) H(v + \eta), \label{BW4}
\end{align}
where $\Theta(v)$ and $H(v)$ are Jacobi theta functions. The necessary relations for the Jacobi's theta and elliptic functions can be found in the book by Baxter~\cite{Bax82}. There are four parameters in all: $v$ --- a spectral parameter, $ \rho $ --- a normalizing factor, $\eta$ --- a parameter, sometimes called the crossing-parameter, and the nome $k$ of theta-functions. We assume that the parameters $\rho$, $\eta$ and $k$ as fixed and omit explicit indication of the dependence on them.

It is useful to have in mind that
\begin{equation}
a(\eta) = c(\eta) = \rho \, \Theta(0) \, H(2 \eta) \, \Theta(2 \eta), \qquad b(\eta) = d(\eta) = 0,
\label{veta}
\end{equation}
and that the following combinations of the weights do not depend on the spectral parameter:
\begin{gather}
\frac{a^2 + b^2 - c^2 - d^2}{a b}=2 \cn (2\eta) \dn (2\eta),  \label{inv1} \\
\frac{c d}{a b}=k \sn^2 (2\eta). \label{inv2}
\end{gather}
Here and below $\sn(v)$, $\cn(v)$ and $\dn(v)$ are Jacobi elliptic functions.

The family of transfer-matrices $\calT(v)$, constructed with the help of the weights (\ref{BW1})--(\ref{BW4}), has the property
\[
[\calT(v), \calT(v')] = 0
\]
for all $v$ and $v'$ \cite{Bax71a, Bax72a}. It means that we can look for vectors $| \Psi \rangle$ which do not depend on $v$ and satisfy the relation
\[
\calT(v) | \Psi \rangle = T(v) | \Psi \rangle,
\]
where $T(v)$ is some function of the spectral parameter. Slightly abusing terminology, we say that $| \Psi \rangle$ is an eigenvector of the transfer-matrix with the eigenvalue $T(v)$.

As was shown by Baxter \cite{Bax72b}, the Hamiltonian of the XYZ-spin chain is closely related to the logarithmic derivative of the transfer-matrix of the eight-vertex model over the spectral parameter $v$ at the point $v = \eta$. Literally repeating his calculations, we obtain
\begin{multline*}
\calT^{-1}(\eta) \, \calT'(\eta) = \frac{N}{2 \, a(\eta)} (a'(\eta) + c'(\eta)) \\*
+ \frac{1}{2 \, a(\eta)} \sum_{j = 1}^N [(b'(\eta) + d'(\eta)) \, \sigma^x_j \sigma^x_{j+1} + (b'(\eta) - d'(\eta)) \, \sigma^y_j \sigma^y_{j+1} + (a'(\eta) - c'(\eta)) \, \sigma^z_j \sigma^z_{j+1}],
\end{multline*}
where prime denotes the derivative over the spectral parameter. Thus,
\begin{equation}
\calT^{-1}  (\eta) \, \calT'(\eta) = \frac{N}{2 \, a(\eta)} (a'(\eta) + c'(\eta)) - \frac{1}{a(\eta)} \calH_{\mathrm{XYZ}}, \label{TH}
\end{equation}
where $\calH_{\mathrm{XYZ}}$ is given by the formula (\ref{H2}) with
\begin{equation}
J_x = b'(\eta) + d'(\eta), \qquad J_y = b'(\eta) - d'(\eta), \qquad J_z = a'(\eta) - c'(\eta). \label{JJJ}
\end{equation}
These relations describe the correspondence between the three fixed parameters of the eight-vertex model and the three parameters of the Hamiltonian of the XYZ-spin chain.

We discuss the simplest symmetry properties of the Hamiltonian $\calH_{\mathrm{XYZ}}$. We use the orthonormal basis, formed by the vectors $| \mu_1 \mu_2 \ldots  \mu_N \rangle$ such that
\[
\sigma^z_j | \mu_1 \mu_2 \ldots \mu_j \ldots \mu_N \rangle = \mu_j | \mu_1 \mu_2 \ldots \mu_j \ldots \mu_N \rangle.
\]
Thus, the quantities $\mu_j$ take values $+1$ and $-1$. If $\mu_j = +1$, we say that the $j$-th spin is up. Similarly, if $\mu_j = -1$, we say that the $j$-th spin is down. It is often convenient to write just $+$ or $-$ instead of $+1$ or $-1$ respectively.

First of all, we note that the Hamiltonian $\calH_{\mathrm{XYZ}}$ is invariant with respect to shifts by one site of the chain to the right or to the left. For definiteness we consider left shifts. The operator $\calS$, implementing this shift can be defined by its action on the basis vectors:
\begin{equation}
\calS | \mu_1 \mu_2 \ldots \mu_N \rangle = | \mu_2 \ldots \mu_N \mu_1 \rangle.
\label{DS}
\end{equation}
The shift invariance of the Hamiltonian $\calH_{\mathrm{XYZ}}$ is expressed by the fact that it commutes with the operator $\calS$.

It is not difficult to get convinced that the Hamiltonian $\calH_{\mathrm{XYZ}}$ is also invariant with respect to rotations about any coordinate axis by the angle $\pi$. First consider rotations about the $z$-axis. The generator of these rotations is
\[
\Sigma^z = \frac{1}{2} \sum_{j = 1}^N \sigma_j^z.
\]
Hence, the operator $\exp(\rmi \, \pi \Sigma^z)$ commutes with the Hamiltonian $\calH_{\mathrm{XYZ}}$. The basis vectors $| \mu_1 \mu_2 \ldots \mu_N \rangle$ are eigenvectors of the operator $N / 2 + \Sigma^z$ with the eigenvalue equal to the number of up spins, and the operator
\begin{equation}
\calP = (-1)^{N/2 + \Sigma^z} = (-1)^N \sigma_1^z  \sigma_2^z \ldots \sigma_N^z
\label{P}
\end{equation}
gives the parity of the number of up spins. Since
\[
\calP = \exp(\rmi \, \pi N / 2)  \exp(\rmi \, \pi \Sigma^z),
\]
the operator $\calP$ also commutes with the Hamiltonian $\calH_{\mathrm{XYZ}}$. Therefore, we can look for eigenvectors of the Hamiltonian restricting ourselves to the sectors with a definite parity of the number of up spins.\footnote{Certainly, as an alternative, we can fix the parity of the number of down spins.}

In a similar way, considering rotations by the angle $\pi$ about the $x$-axis, we conclude that the operator
\begin{equation}
\calI = \sigma_1^x \sigma_2^x \ldots \sigma_N^x, \label{I}
\end{equation}
inverting the direction of all spins, commutes with the Hamiltonian of the chain. Hence, if $| \Psi \rangle$ is an eigenvector of the Hamiltonian $\calH_{\mathrm{XYZ}}$, then the vector $\calI | \psi \rangle$ is also an eigenvector with the same eigenvalue. In particular, if the length of the chain is odd, then any eigenvalue is, at least doubly, degenerate.

Rotations by the angle $\pi$ about the $y$-axis can be represented as combinations of rotations by the angle $\pi$ about the $x$-axis and the $z$-axis. Therefore, consideration of these rotations do not lead to additional conclusions on eigenvectors and eigenvalues of the Hamiltonian $\calH_{\mathrm{XYZ}}$.

\section{Simple eigenvalue}

In this section, using the connection with the eight-vertex model, we give arguments supporting that in the case when the condition (\ref{condition}) is satisfied and $N$ is odd, the Hamiltonian $\calH_{\mathrm{XYZ}}$ has an eigenvector with the eigenvalue (\ref{sev}).

We rewrite the condition (\ref{condition}) in terms of the fixed parameters of the eight-vertex model. To this end, we first rewrite the equalities (\ref{inv1}) and (\ref{inv2}) in a slightly modified form:
\begin{gather*}
a^2 + b^2 - c^2 - d^2 = 2 \cn (2\eta) \dn (2\eta) \, a  b,  \nonumber \\
c d=k \sn^2(2\eta) \, a b, \nonumber
\end{gather*}
differentiate them over the spectral parameter $v$ and put $v=\eta$. Taking into account the equalities (\ref{veta}), we obtain
\begin{gather}
a'(\eta) - c'(\eta) = \cn (2\eta)  \dn (2\eta) \, b'(\eta),  \label{inv3a} \\
d'(\eta) = k  \sn^2 (2\eta)  \, b'(\eta).  \label{inv3b}
\end{gather}
Hence, we can write the relations (\ref{JJJ}) in the form
\[
J_x  = [1 + k \sn^2(2\eta)] b'(\eta), \qquad J_y  = [1 - k \sn^2(2\eta)] b'(\eta), \qquad J_z = \cn (2\eta) \dn (2\eta) b'(\eta)
\]
and see that the equality
\[
J_x J_y + J_y J_z + J_z J_x = [1 - k^2 \sn^4 (2 \eta) + 2 \cn (2\eta) \dn (2\eta)] b'^2(\eta)
\]
is valid. Thus, the quantities  $J_x$, $J_y$ and $J_z$ satisfy the relation (\ref{condition}) either if $b'(\eta) = 0$ or if
\begin{equation}
1 - k^2 \sn^4 (2 \eta) + 2 \cn (2\eta) \dn (2\eta) = 0.
\label{condition1}
\end{equation}
In the case when $b'(\eta) = 0$, we obtain trivial values of $J_x$, $J_y$ and $J_z$. Assuming that $\eta$ is real, one can show up to periodicity that the equality (\ref{condition1}) is valid only at $\eta = \pm 2 K / 3$ and at $\eta = \pm 4 K / 3$, where $K$ is the complete elliptic integral of the first kind. Here the obtained values of $J_x$, $J_y$ and $J_z$ do not depend on the choice of a possible value of $\eta$.

Now we consider the inversion relations which should be satisfied by the eigenvalues of the transfer-matrix~\cite{Bax80, Str79}:
\begin{equation}
\label{Inversion}
T(v - \eta)\, T(v + \eta) = \phi^N(v - 2\eta) \phi^N(v + 2\eta) + \phi^N(v) P(v).
\end{equation}
Here the eigenvalue of the transfer-matrix $T(v)$ and the function $P(v)$ are some quasi-periodic entire functions of order $N$, and
\begin{equation*}
\phi (v) = \rho \, \Theta (0) \, H(v) \, \Theta (v).
\end{equation*}
We stress again that an arbitrary solution to the inversion relations satisfying the necessary quasi-periodicity requirements can be not an eigenvalue of the transfer-matrix due to absence of the corresponding eigenvector.

In the case when  $\eta = \pm 2K / 3$ or $\eta = \pm 4 K / 3$, there are simple solutions to the inversion relations:
\begin{equation}
\label{T}
T(v) = \phi^N(v), \qquad P(v) = 0
\end{equation}
and
\begin{equation}
T(v) = - \phi^N(v), \qquad P(v) = 0.
\label{T1}
\end{equation}
To get convinced in this, it is enough to use the properties
\begin{equation*}
\Theta(v \pm 2 K) = \Theta(v), \qquad H(v \pm 2 K) = - H(v),
\end{equation*}
which imply
\begin{equation*}
\phi (v \pm 2 K) = - \phi(v).
\end{equation*}
Using this equality, it is not difficult to get convinced that (\ref{T}) and (\ref{T1}) are really solutions to inversion relations (\ref{Inversion}).

It is convenient to have an expression for $T(v)$ in terms of the Boltzmann weights (\ref{BW1})--(\ref{BW4}). Consider the sum
\begin{equation*}
a(v) + b(v)  = \rho \, \Theta (2\eta) [ \Theta (v - \eta) H(v + \eta) + \Theta(v + \eta) H(v - \eta)].
\end{equation*}
Dividing the right hand side of this equality by $\phi(v)$, we obtain an elliptic function without poles, i. e., a constant. Assuming that $v = 2 \eta$, we see that this constant has the form
\begin{equation*}
\frac{\Theta(\eta) H(3 \eta) + \Theta(3 \eta) H(\eta)}{\Theta(0) H(2 \eta)},
\end{equation*}
that in the cases $\eta = 2 K/3$ and $\eta = - 4 K/3$ is equal to $1$, and in the cases $\eta = - 2 K/3$ and $\eta = 4 K/3$ is equal to $-1$. Hence,
\begin{equation}
   \label{Tab}
T(v) = [a(v) + b(v)]^N,
\end{equation}
if for $\eta = 2 K/3$ and $\eta = - 4 K/3$ we take the solution (\ref{T}), and for $\eta = - 2 K/3$ and $\eta = 4 K/3$ the solution (\ref{T1}), and
\begin{equation*}
T(v) = (-1)^N [a(v) + b(v)]^N,
\end{equation*}
if for $\eta = 2 K/3$ and $\eta = - 4 K/3$ we take the solution (\ref{T1}), and for $\eta = - 2 K/3$ and $\eta = 4 K/3$ the solution (\ref{T}). It is possible to verify that for small values of $N$ only (\ref{Tab}) and only for odd $N$ is an eigenvalue of the transfer-matrix.

\begin{conjecture} \label{hc}
At $\eta = \pm 2 K/3$ and at $\eta = \pm 4 K/3$ the transfer-matrix $\calT(v)$ of the eight-vertex model for the case of periodic boundary conditions and an odd number of sites $N$ in the horizontal direction has an eigenvector with the eigenvalue
\[
T(v) = [a(v) + b(v)]^N.
\]
\end{conjecture}

As follows from the equality (\ref{TH}), the eigenvalue (\ref{Tab}) of the transfer-matrix  corresponds to the eigenvalue $- N(a'(\eta) + 2 b'(\eta) - c'(\eta) / 2$ of the Hamiltonian $\calH_{\mathrm{XYZ}}$. Using the equalities (\ref{JJJ}), it is not difficult to get convinced that this eigenvalues just coincides with $-N(J_x + J_y + J_z) / 2$. We fix the energy scale to satisfy the equality
\[
J_x + J_y = 2,
\]
that is consistent with the fixation of the energy scale for the XXZ-spin chain. Introducing the notation
\begin{equation}
\alpha = k \sn^2 (2\eta)
\label{alpha}
\end{equation}
and taking into account the relation (\ref{condition1}), we obtain
\begin{equation}
\label{JxJy}
J_x = 1 + \alpha, \qquad J_y = 1 - \alpha, \qquad J_z  = (\alpha ^2 - 1) / 2.
\end{equation}
It follows from here that the eigenvalue under consideration has the form
\begin{equation*}
E(\alpha) = - \frac{N}{2} (J_x + J_y + J_z)= - \frac{N}{4} (3 + \alpha^2).
\end{equation*}
We denote the Hamiltonian $\calH_{\mathrm{XYZ}}$ with $J_x$, $J_y$ and $J_z$, given by the equalities (\ref{JxJy}), by $\calH(\alpha)$ and formulate the following conjecture.

\begin{conjecture} \label{c:3.2}
For an odd number of sites $N$ the ground state of the Hamiltonian $\calH(\alpha)$ is doubly degenerate and has the energy
\begin{equation}
E(\alpha) =  - \frac{N}{4} (3 + \alpha^2). \label{in}
\end{equation}
\end{conjecture}

It is worth to note a few special values of the parameter $\alpha$. At $\alpha=0$ we have XXZ-spin chain with $J_x = J_y = 1$ and $J_z = - 1/2$. At $\alpha=\pm 3$ we obtain the XXZ-spin chain with $J_x = J_y = 1$ and $J_z = - 1/2$, rotated by the angle $\pi / 2$ about the $x$-axis or the $y$-axis. In the case $\alpha=1$ we obtain a trivial Hamiltonian $\calH(0) = - \sum_{j=1}^{N} \sigma_j^x \sigma_{j+1}^x$.

In the case when  $\eta = \pm 2K / 3$ or $\eta = \pm 4 K / 3$ the inversion relations of the inhomogeneous eight-vertex model
\begin{multline*}
T(v - \eta | v_1, \ldots, v_N) \, T(v + \eta | v_1, \ldots, v_N) \\ = \prod_{j=1}^N [ \phi(v - v_j - 2\eta) \phi(v - v_j + 2 \eta)] + \left[ \prod_{j=1}^N \phi(v - v_j) \right] P(v | v_1, \ldots, v_N).
\end{multline*}
also have two simple solutions:
\begin{equation*}
T(v | v_1, \ldots, v_N) = \prod_{j=1}^N \phi(v - v_j), \qquad P(v | v_1, \ldots, v_N) = 0
\end{equation*}
and
\begin{equation*}
T(v | v_1, \ldots, v_N) = -\prod_{j=1}^N \phi(v - v_j), \qquad P(v | v_1, \ldots, v_N) = 0.
\end{equation*}
Using argumentation similar to used above, we conclude that the following congecture is very plausible.

\begin{conjecture} \label{uc}
At $\eta = \pm 2 K/3$ or at $\eta = \pm 4 K/3$ the inhomogeneous transfer-matrix $\calT(v | v_1, \ldots, v_N)$ of the eight-vertex model for the case of the periodic boundary conditions and an odd number of sites $N$ in the horizontal direction has an eigenvector with the eigenvalue
\begin{equation}
T(v | v_1, \ldots, v_N) = \prod_{j=1}^N [a(v - v_j) + b(v - v_j)].
\label{uev}
\end{equation}
\end{conjecture}

As in the homogeneous case, since the transfer-matrices $\calT(v | v_1, \ldots, v_N)$ with different values of the spectral parameter $v$ commute, we can assume that the eigenvector of the Corollary \ref{uc} does not depend on $v$.

Concluding this section, note that, using the equalities (\ref{inv1}) and (\ref{inv2}), we can write (\ref{condition1}) as an elegant condition on the weights
\begin{equation*}
 (a^2 + a b) (b^2 + a b) = (c^2 + a b) (d^2 + a b),
\end{equation*}
which we will not use however.

\section{Conjectures on properties of components}

The matrix elements of the Hamiltonian $\calH(\alpha)$ with respect to the basis under consideration as well as the eigenvalue $E(\alpha)$ are polynomials in $\alpha$ with rational coefficients. On can normalize an eigenvector $| \Psi(\alpha) \rangle$, corresponding to the eigenvalue $E(\alpha)$, so that all its components are polynomials in $\alpha$ with integer coefficients.

In the limit of the XXZ-spin chain, when $\alpha = 0$, the components are simply integers, and we noticed~\cite{RazStr01a} that they are {\em positive\/} integers and related to enumerations of the alternating-sign matrices.

In this and the next sections we study the properties of the components for the XYZ-spin chain and see, in particular, that in this case there are indications of a possible relation to combinatorial problems as well.

For illustrative purposes we give the explicit expressions for the components of the vector $| \Psi(\alpha) \rangle$ for $N = 3$, $5$ and $7$. For each odd $N$ there are two eigenvectors with the eigenvalue $E(\alpha)$, which differ by the parity of the number of up spins. We will use the notation $| \Psi(\alpha) \rangle$ for the eigenvector belonging to the sector with positive parity of the number of up spins. The corresponding vector belonging to the sector with negative parity of the number of up spins will be denoted $| \bar{\Psi}(\alpha) \rangle$. In accordance with our convention
\[
\calP | \Psi(\alpha) \rangle = | \Psi(\alpha) \rangle, \qquad \calP | \bar \Psi(\alpha) \rangle = - | \bar \Psi(\alpha) \rangle,
\]
where the operator $\calP$ is defined by the equality (\ref{P}), and we can assume that $| \bar \Psi(\alpha) \rangle = \calI | \Psi(\alpha) \rangle$, where the operator $\calI$ is defined by the equality (\ref{I}). The components of the vectors $| \Psi(\alpha) \rangle$ and $| \bar \Psi(\alpha) \rangle$ are connected by the evident relation
\begin{equation}
\bar \Psi_{\mu_1 \mu_2 \ldots \mu_n}(\alpha) = \Psi_{\bar \mu_1 \bar \mu_2 \ldots \bar \mu_n}(\alpha),
\label{tobar}
\end{equation}
we we denote $\bar \mu_j = - \mu_j$. Therefore, we give expressions only for the components of the vector $| \Psi(\alpha) \rangle$.

For the case $N=3$ the nonzero components of the vector $| \Psi(\alpha) \rangle$ are
\begin{equation*}
\Psi_{---}(\alpha) = \alpha, \qquad \Psi_{-++}(\alpha) = \Psi_{+-+}(\alpha) = \Psi_{++-}(\alpha) = 1.
\end{equation*}
Note that for $N = 3$ the vector $| \Psi(\alpha) \rangle$, as well as the vector $| \bar \Psi(\alpha) \rangle$, is shift invariant. This is valid for all odd $N$, at least, up to $N = 19$.

\begin{conjecture}
The vectors $| \Psi(\alpha) \rangle$ and $| \bar \Psi(\alpha) \rangle$ are shift invariant, i. e., the equality
\[
\calS | \Psi(\alpha) \rangle = | \Psi(\alpha) \rangle, \qquad \calS | \bar \Psi(\alpha) \rangle = | \bar \Psi(\alpha) \rangle,
\]
where the operator $\calS$ is defined by the equality {\rm (\ref{DS})}, is valid.
\end{conjecture}

We give now the expressions for the nonzero components of the vectors $| \Psi(\alpha) \rangle$ for $N = 5$ and $N = 7$, which cannot be obtained one from another by a shift of indices:
\begin{align*}
\boxed{N = 5} \hskip -1.em \\*[.5em]
&\Psi_{-----}(\alpha) = \alpha + \alpha^3,
&& \Psi_{---++}(\alpha) = 1 + \alpha^2, \\
& \Psi_{--+-+}(\alpha) = 2,
&& \Psi_{-++++}(\alpha) = 2 \alpha; \\[.5em]
\boxed{N = 7} \hskip -1.em \\*[.5em]
& \Psi_{-------}(\alpha) = 4 \alpha^2 + 3 \alpha^4 + \alpha^6,
&& \Psi_{-----++}(\alpha) = 4 \alpha + 3 \alpha^3 +\alpha^5, \\
& \Psi_{----+-+}(\alpha) = 7 \alpha + \alpha^3,
&& \Psi_{---+--+}(\alpha) = 7 \alpha + \alpha^3, \\
& \Psi_{---++++}(\alpha) = 1 + 5 \alpha^2 + 2 \alpha^4,
&& \Psi_{--+-+++}(\alpha) = 3 + 5 \alpha^2, \\
& \Psi_{--++-++}(\alpha) = 4 + 3 \alpha^2 + \alpha^4,
&& \Psi_{-+--+++}(\alpha) = 3 + 5 \alpha^2, \\
& \Psi_{-+-+-++}(\alpha) = 7 + \alpha ^2,
&& \Psi_{-++++++}(\alpha) = 3 \alpha + 5 \alpha^3.
\end{align*}
We use the normalization under which the components are polynomial in $\alpha$ with minimally possible integer coefficients. One can notice immediately that all components are polynomials with {\em positive\/} coefficients that argues in favor of a possible combinatorial interpretation.

Considering the limit of the relation
\begin{equation}
\calH(\alpha) | \Psi(\alpha) \rangle = E(\alpha) | \Psi(\alpha) \rangle
\label{HPsi}
\end{equation}
as $\alpha$ tends to infinity, it is easy to understand that among the components of the vector $| \Psi(\alpha) \rangle$ the component $\Psi_{-- \cdots -}(\alpha)$ is a polynomial in $\alpha$ of the maximal degree. We denote this degree by $D_N$. One can notice, that under the used normalization the coefficient at $\alpha^{D_N}$ in $\Psi_{-- \cdots -}(\alpha)$ is equal to $1$.

\begin{conjecture}
For $N = 2n + 1$ the degree of the polynomial $\Psi_{-- \cdots -}(\alpha)$ is given by the formula
\[
D_N = (N^2 - 1) / 8 = n(n + 1)/2.
\]
\end{conjecture}

\begin{conjecture} \label{c:5}
We normalize the vector $| \Psi(\alpha) \rangle$ so that
\[
\Psi_{-- \cdots -}(\alpha) = \alpha^{D_N} + o(\alpha^{D_N}).
\]
Under such normalization the components of the vector $| \Psi(\alpha) \rangle$ are polynomials in $\alpha$  with positive integer coefficients.

In the limit as $\alpha$ tends to $0$ we obtain a ground state vector of the XYZ-spin chain under the normalization where the minimal component of the vector is equal to $1$.
\end{conjecture}

\begin{conjecture}
Under the normalization of Conjecture \ref{c:5}, we have for $N = 4m + 1$
\[
\Psi_{-- \cdots -}(\alpha) = \alpha^{(2m + 1)m} + \cdots + A_{\mathrm V}(2m + 1)^2 \alpha^m,
\]
where $A_V(2m + 1)$ is the number of vertically symmetric alternating-sign matrices of order $2m + 1$, and for $N = 4m - 1$
\[
\Psi_{-- \cdots -}(\alpha) = \alpha^{(2m - 1)m} + \cdots + N_8(2m)^2 \alpha^m,
\]
where $N_8(2m)$ is the number of cyclically symmetric transpose complement plane partitions fitting inside a box of size $2m \times 2m \times 2m$.
\end{conjecture}

A formula for $A_V(2m + 1)$ was conjectured by Robbins \cite{Rob00} and proved by Kuperberg \cite{Kup02}. We give it in the form borrowed from our paper \cite{RazStr04bE}:
\[
A_{\mathrm V}(2m + 1) = \frac{1}{2^m} \prod_{i=0}^{m-1} \frac{(6i + 4)! (2i + 1)!}{(4i + 3)! (4i + 2)!}.
\]
A formula for $N_8(2m)$, having the form
\[
N_8(2m) = \prod_{i=0}^{m-1} (3i + 1) \frac{(6i)! (2i)!}{(4i + 1)! (4i)!},
\]
was proved in the paper by Mills, Robbins and Rumsey \cite{MilRobRum83}.

As we already noted, rotations about coordinate axes by the angle $\pi$ are symmetry transformations for the Hamiltonian $\calH_{\mathrm{XYZ}}$, and, therefore, for the operator $\calH(\alpha)$. It appears that consideration of rotations by the angle $\pi/2$ above the coordinates axes, which are not symmetry transformations, allow to obtain useful information on the vectors $| \Psi(\alpha) \rangle$ and $| \bar \Psi(\alpha) \rangle$.

First, we consider rotations by the angle $\pi/2$ about the $z$-axis. The operator
\[
\rho^z (\pi/2) = \cos (\pi / 4) + \rmi \, \sigma^z \sin (\pi / 4) = (1 + \rmi \, \sigma^z)/\sqrt 2
\]
describes rotations of an individual spin, and rotations of the whole chain are described by the operator
\[
\calR^z(\pi/2) = \rho^z_1(\pi/2) \, \rho^z_2(\pi/2) \ldots \rho^z_N(\pi/2).
\]
Using the relations
\begin{gather*}
\calR^z(\pi/2) \, \sigma^x_j = - \sigma^y_j \, \calR^z(\pi/2), \qquad \calR^z(\pi/2) \, \sigma^y_j = \sigma^x_j \, \calR^z(\pi/2), \\[.5em]
\calR^z(\pi/2) \, \sigma^z_j = - \sigma^z_j \, \calR^z(\pi/2),
\end{gather*}
it is not difficult to show that the equality
\begin{equation}
\calR^z(\pi/2) \, \calH(\alpha) = \calH(-\alpha) \, \calR^z(\pi/2)
\label{RH}
\end{equation}
is valid.

We act on the both sides of the equality (\ref{HPsi}) by the operator $\calR^z(\pi/2)$. Taking into account the relation (\ref{RH}) and changing $\alpha$ to $-\alpha$, we obtain
\[
\calH(\alpha) \left[ \calR^z(\pi/2) | \Psi(-\alpha) \rangle \right] = E(\alpha) \left[ \calR^z(\pi/2) | \Psi(-\alpha) \rangle \right].
\]
If Conjecture \ref{c:3.2} is valid, the eigenvalue $E(\alpha)$ in a sector with a definite parity of the number of up spins is nondegenerate. Taking into account that the operator $\calR^z(\pi/2)$ does not change the parity of the number of up spins, we conclude that the vector $\calR^z(\pi/2) | \Psi(-\alpha) \rangle$ is proportional to the vector $| \Psi(\alpha) \rangle$. It is not difficult to understand that this is possible only in the case when the components of $| \Psi(\alpha) \rangle$ are polynomials in $\alpha$ of definite parity. We formulate this statement in a more concrete form as a corollary of our conjectures.

\begin{corollary}
For any odd $N$ we have
\begin{multline*}
\Psi_{\mu_1 \ldots \mu_N}(-\alpha) =  (-1)^{D_N} (-1)^{ (N + \mu_1 + \ldots + \mu_n) / 4 } \Psi_{\mu_1 \ldots \mu_N}(\alpha) \\
= (-1)^{(N^2 - 1) / 8} (-1)^{ (N + \mu_1 + \ldots + \mu_n) / 4 } \Psi_{\mu_1 \ldots \mu_N}(\alpha).
\end{multline*}
\end{corollary}

Considering rotations by the angle $\pi / 2$ about the $x$-axis and the $y$-axis, we obtain
\begin{gather}
\calR^x(\pi/4) \calH(\alpha) = \frac{(1 + \alpha)^2}{4} \calH((3 - \alpha)/(1 + \alpha)) \calR^x(\pi/4),
\label{RH1} \\
\calR^y(\pi/4) \calH(\alpha) = \frac{(\alpha - 1)^2}{4} \calH((\alpha + 3)/(\alpha - 1)) \calR^y(\pi/4).
\label{RH2}
\end{gather}
These equalities describe, in particular, the connection of the eigenvectors of the Hamiltonian of the XXZ-spin chain, corresponding to the case $\alpha = 0$, and the eigenvectors of the Hamiltonians of the rotated XXZ-spin chains, corresponding to the cases $\alpha = 3$ and $\alpha = -3$.

\section{Sum rules for components}

In this section we consider the linear and quadratic sums of the components of the vectors $| \Psi(\alpha) \rangle$ and $| \bar \Psi(\alpha) \rangle$. Before all, we introduce the notation
\[
S_1(\alpha) = \sum_{\mu_1, \ldots, \mu_N} \Psi_{\mu_1 \ldots \mu_N}(\alpha), \qquad S_2(\alpha) = \sum_{\mu_1, \ldots, \mu_N} \Psi^2_{\mu_1 \ldots \mu_N}(\alpha)
\]
and note that it follows from the relation (\ref{tobar}) that
\[
\sum_{\mu_1, \ldots, \mu_N} \bar \Psi_{\mu_1 \ldots \mu_N}(\alpha) = S_1(\alpha), \qquad  \sum_{\mu_1, \ldots, \mu_N} \bar \Psi^2_{\mu_1 \ldots \mu_N}(\alpha) = S_2(\alpha).
\]
The first our observation on the properties of $S_1(\alpha)$ and $S_2(\alpha)$ is the fact that the polynomial $S_1(\alpha)$ almost divides the polynomial $S_2(\alpha)$.

\begin{conjecture}
If $N = 4m - 1$ or $N = 4m + 1$ then
\[
\frac{S_2(\alpha)}{S_1(\alpha)} = \frac{F(\alpha)}{(\alpha + 3)^m},
\]
where $F(\alpha)$ is a polynomial in $\alpha$.
\end{conjecture}

We consider again rotations by the angle $\pi / 2$ about the $y$-axis and denote
\[
| \Phi(\alpha) \rangle = \calR^y(\pi / 2) | \Psi(\alpha) \rangle.
\]
As follows from the relation (\ref{RH2}), the vector $| \Phi(\alpha) \rangle$ is an eigenvector of the Hamiltonian $\calH(\alpha')$ with the eigenvalue $E(\alpha')$, where
\[
\alpha' = \frac{\alpha + 3}{\alpha - 1}.
\]
Assume that Conjecture \ref{c:3.2} is valid, then $| \Phi(\alpha) \rangle$ is a linear combination of the vectors $| \Psi(\alpha') \rangle$ and $|\bar  \Psi(\alpha') \rangle$. Since
\[
\rho^y (\pi / 2) = \cos (\pi / 4) + \rmi \, \sigma^y \sin (\pi / 4) = (1 + \rmi \, \sigma^z)/\sqrt 2,
\]
we have
\begin{equation}
\rho^y(\pi / 2) | + \rangle = \frac{1}{\sqrt{2}}  | + \rangle - \frac{1}{\sqrt{2}} | -  \rangle, \qquad \rho^y(\pi / 2) | - \rangle = \frac{1}{\sqrt{2}} | + \rangle +  \frac{1}{\sqrt{2}} | - \rangle,
\label{rhop}
\end{equation}
therefore
\[
\calR^y(\pi / 2) | \mu_1 \ldots \mu_N \rangle = \frac{1}{\sqrt{2^N}} | + \ldots + \rangle + \ldots + (-1)^{(N + \mu_1 + \ldots + \mu_N) / 2} \frac{1}{\sqrt{2^N}} | - \ldots - \rangle.
\]
Using this relation, it is not difficult to demonstrate that
\begin{equation}
| \Phi(\alpha) \rangle = \frac{S_1(\alpha)}{\sqrt{2^N}}  \left[ \frac{1}{\Psi_{- \cdots -}(\alpha')} | \Psi(\alpha') \rangle + \frac{1}{\bar \Psi_{+ \cdots +}(\alpha')} | \bar \Psi(\alpha') \rangle \right].
\label{phivec}
\end{equation}

It follows from the equalities  (\ref{rhop}) that if $| \varphi  \rangle = \rho^y(\pi / 2) | \psi \rangle$, then
\[
\sum_\mu \varphi_\mu = \sqrt 2 \, \psi_-, \qquad  \sum_\mu \psi_\mu = \sqrt 2 \, \varphi_+.
\]
Now one can get convinced that
\begin{equation}
\sum_{\mu_1, \ldots, \mu_N} \Phi_{\mu_1 \ldots \mu_N} (\alpha) = \sqrt{2^N} \Psi_{- \cdots -}(\alpha), \qquad \sum_{\mu_1, \ldots, \mu_N} \Psi_{\mu_1 \ldots \mu_N} (\alpha) = \sqrt{2^N} \Phi_{+ \cdots +}(\alpha).
\label{sphi}
\end{equation}

We rewrite the equality (\ref{phivec}) in terms of the components:
\begin{equation}
\Phi_{\mu_1 \ldots \mu_N}(\alpha) = \frac{S_1(\alpha)}{\sqrt{2^N}}  \left[ \frac{\Psi_{\mu_1 \ldots \mu_N}(\alpha')}{\Psi_{- \cdots -}(\alpha')} + \frac{\bar \Psi_{\mu_1 \ldots \mu_N}(\alpha')}{\bar \Psi_{+ \cdots +}(\alpha')} \right].
\label{phicomp}
\end{equation}
Summing over the indices and taking into account the first equality of (\ref{sphi}) and the relation (\ref{tobar}), we obtain the following corollary of our conjectures.

\begin{corollary}
The relation
\begin{equation}
\frac{S_1(\alpha)}{\Psi_{- \cdots -}(\alpha)} \, \frac{S_1(\alpha')}{\Psi_{- \cdots -}(\alpha')} = 2^{N-1},
\label{S1S1}
\end{equation}
where $\alpha' = (\alpha + 3) / (\alpha - 1)$, is valid.
\end{corollary}

Using the explicit form of the components of the vectors $| \Psi(\alpha) \rangle$ for odd $N \le 19$, we conclude that apparently the following congecture is valid.

\begin{conjecture} \label{c:5.1}
The equalities
\begin{gather*}
S_1(\alpha) = 2^{-(N-3)(N-1) / 8} \, (\alpha - 1)^{D_N} \Psi_{- \cdots -}(\alpha'), \\
S_1(\alpha') = 2^{(N-1)(N+5) / 8} \, (\alpha - 1)^{-D_N} \Psi_{- \cdots -}(\alpha),
\end{gather*}
where $\alpha' = (\alpha + 3) / (\alpha - 1)$, are valid.
\end{conjecture}

The matrix $\rho^y(\pi / 2)$ is orthogonal, hence,
\begin{equation}
\sum_{\mu_1, \ldots, \mu_N} \Phi_{\mu_1 \ldots \mu_N}^2(\alpha) = \sum_{\mu_1, \ldots, \mu_N} \Psi_{\mu_1 \ldots \mu_N}^2(\alpha).
\label{phipsi}
\end{equation}
Using the relations (\ref{phipsi}) and (\ref{phicomp}), together with the fact that the components $\Psi_{\mu_1 \ldots \mu_N}(\alpha)$ and $\bar \Psi_{\mu_1 \ldots \mu_N}(\alpha)$ are nonzero for different sets of indices, we obtain
\[
2^{N-1} \Psi_{- \cdots -}^2(\alpha') \, S_2(\alpha) = S_1^2(\alpha) \, S_2(\alpha').
\]
Proceeding from this and taking into account the equality (\ref{S1S1}), we formulate one more corollary.

\begin{corollary}
The equality
\[
\frac{S_2(\alpha)}{S_1(\alpha) \Psi_{- \cdots -}(\alpha)} = \frac{S_2(\alpha')}{S_1(\alpha') \Psi_{- \cdots -}(\alpha')},
\]
where $\alpha' = (\alpha + 3) / (\alpha - 1)$, is valid.
\end{corollary}

If Conjecture \ref{c:5.1} is valid, this corollary can be reformulated as follows.

\begin{corollary} \label{cr:5.3}
The equality
\[
S_2(\alpha') = 2^{2 D_N} (\alpha - 1)^{-2 D_N} S_2(\alpha),
\]
where $\alpha' = (\alpha + 3) / (\alpha - 1)$, is valid.
\end{corollary}

Introducing the notation
\[
\tilde S_2(\alpha) = (\alpha + 1)^{-2 D_N} S_2(\alpha),
\]
one can formulate the statement of Corollary \ref{cr:5.3} as the statement that the rational function $\tilde S_2(\alpha)$ is invariant with respect to the linear fractional transformation $\alpha  \to (\alpha + 3) / (\alpha - 1)$.

\section{Chain degree of antiferromagneticity}

We consider the transfer-matrix $\calT(v | v_1, \ldots, v_N)$ of the inhomogeneous eight-vertex model. At $v_j = 0$ the transfer-matrix $\calT(v | v_1, \ldots, v_N)$ is reduced to the transfer-matrix of the homogeneous model and its eigenvectors to eigenvectors of the homogeneous model, which are also eigenvectors of the Hamiltonian $\calH_{\mathrm{XYZ}}$.

Now we assume that $\eta = \pm 2 K / 3$ or $\eta = \pm 4 K / 3$ and that Conjecture \ref{uc} is valid. We perform the transition to the homogeneous case in two steps. First we put $v_1 = 0$ and $v_j = v - \eta$, $j = 2, \ldots, N$, and then put $v = \eta$. Let $| \Psi(v_1, \ldots, v_N) \rangle$ be an eigenvector of the transfer-matrix with the eigenvalue given by the relation (\ref{uev}), i. e., the equality
\[
\calT(v | v_1, \ldots, v_N) | \Psi(v_1, \ldots, v_N) \rangle = \left[ \prod_{j=1}^N [a(v - v_j) + b(v - v_j)] \right]  | \Psi(v_1, \ldots, v_N) \rangle
\]
is valid. Assuming that in this equality $v_1 = 0$ and $v_j = v - \eta$, $j = 2, \ldots, N$, and having in mind that $b(\eta) = 0$, we obtain
\begin{equation}
\calU(v) | \Phi(v) \rangle = [a(v) + b(v)]  | \Phi(v) \rangle,
\label{UPhi}
\end{equation}
where $| \Phi(v) \rangle = | \Psi(0, v - \eta, \ldots, v - \eta) \rangle$, and
\[
\calU(v) =\frac{1}{ a^{N-1}(\eta)} \calT(v | 0, v - \eta, \ldots, v - \eta).
\]
The operator $\calU(v)$ at $v = \eta$ coincides with the shift operator defined by the equality (\ref{DS}), while for an arbitrary $v$ it is a shift operator with one defect. Using the standard definition of the transfer-matrix, we see that the equation (\ref{UPhi}) is equivalent to the equations
\begin{gather}
a(v) \Phi_{\mu \mu \mu_3 \ldots \mu_N}(v) + d(v) \Phi_{\bar \mu \bar \mu \mu_3 \ldots \mu_N}(v) = [a(v) + b(v)] \Phi_{\mu \mu_3 \ldots \mu_N \mu}(v), \label{milla} \\[.5em]
c(v) \Phi_{\mu \bar \mu \mu_3 \ldots \mu_N}(v) + b(v) \Phi_{\bar \mu \mu \mu_3 \ldots \mu_N}(v) = [a(v) + b(v)] \Phi_{\mu \mu_3 \ldots \mu_N \bar \mu}(v). \label{millb}
\end{gather}
Performing in the equalities (\ref{milla}) and (\ref{millb}) the summation over the indices and adding the obtained equalities, we obtain
\begin{multline}
[a(v) + d(v)] \sum_{\mu, \mu_3, \ldots, \mu_N}   \Phi_{\mu \mu \mu_3 \ldots \mu_N}(v) + [c(v) + b(v)] \sum_{\mu, \mu_3, \ldots, \mu_N}   \Phi_{\mu \bar \mu \mu_3 \ldots \mu_N}(v) \\
= [a(v) + b(v)] \sum_{\mu_1, \mu_2, \mu_3, \ldots, \mu_N}   \Phi_{\mu_1 \mu_2 \mu_3  \ldots \mu_N}(v). \label{le}
\end{multline}
Taking into account the identity
\[
\sum_{\mu_1, \mu_2, \mu_3, \ldots, \mu_N}   \Phi_{\mu_1 \mu_2 \mu_3  \ldots \mu_N}(v) = \sum_{\mu, \mu_3, \ldots, \mu_N}   \Phi_{\mu \mu \mu_3  \ldots \mu_N}(v) + \sum_{\mu,  \mu_3, \ldots, \mu_N}   \Phi_{\mu \bar \mu_2 \mu_3  \ldots \mu_N}(v),
\]
we write the equality (\ref{le}) in the form
\begin{equation*}
\frac{\sum_{\mu, \mu_3, \ldots, \mu_N}   \Phi_{\mu \bar \mu \mu_3 \ldots \mu_N}(v)}{\sum_{\mu, \mu_3, \ldots, \mu_N}   \Phi_{\mu \mu \mu_3 \ldots \mu_N}(v)} = \frac{d(v) - b(v)}{a(v) - c(v)}.
\end{equation*}
When $v$ tends to $\eta$, the vector $ | \Phi(v) \rangle$ tends to the vector $ | \Psi(\alpha) \rangle$. Using the l'H\^opital's rule, the relations (\ref{inv3a})--(\ref{condition1}) and the notation (\ref{alpha}), we obtain the last, for this paper, corollary of our conjectures.

\begin{corollary}
\begin{equation*}
\Xi(\alpha) = \frac{\sum_{\mu, \mu_3, \ldots, \mu_N}   \Psi_{\mu \bar \mu \mu_3 \ldots \mu_N}(\alpha)}{\sum_{\mu, \mu_3, \ldots, \mu_N}   \Psi_{\mu \mu \mu_3 \ldots \mu_N}(\alpha)} = \frac{2}{\alpha + 1}.
\end{equation*}
\end{corollary}

The quantity $\Xi(\alpha)$ can be considered as an estimation of the chain degree of antiferromagneticity.

\section{Conclusion}

In the present paper we presented first results of studying the eigenvector of the Hamiltonian of the XYZ-spin chain, corresponding to a simple eigenvalue, which exists under a special choice of the parameters of the Hamiltonian. It seems very plausible that this vector is a ground state vector that is confirmed by the results obtained in the limit of the XXZ-spin chain. Under an appropriate normalization, the components of the vector under consideration with respect to the natural basis are polynomials with positive integer coefficients in the only remaining free parameter of the Hamiltonian. This allows us to expect some connection with combinatorial problems. A connection to enumerations of plain partitions already became apparent.

{\em Acknowledgments\/}. We are thankful to P. Di Francesco, V. Pasquier and P. Zinn-Justin for interesting and useful discussions, and to V. Bazhanov and V. Mangazeev for a substantial correspondence. The work was supported in part by RFBR grants 07--01--00234 and 09--01--93107. A.V.R. was also supported in part by RFBR grant 09-01-12123.

\newcommand{\noopsort}[1]{}
\providecommand{\href}[2]{{#2}}
\providecommand{\curlanguage}[1]{%
 \expandafter\ifx\csname #1\endcsname\relax
 \else\csname #1\endcsname\fi}

\end{document}